\documentclass[twocolumn,showpacs,preprintnumbers,prl,aps,amssymb]{revtex4}
\usepackage{graphicx}
\usepackage{dcolumn}
\usepackage{bm}

\begin{document}

\title{Andreev reflection spectroscopy on the heavy-fermion superconductor PrOs$_{4}$Sb$_{12}$: Microscopic evidence for multiple superconducting order parameters}

\author{C. S.~Turel$^1$, W. M.~Yuhasz$^2$, R.~Baumbach$^2$, M. B.~Maple$^2$ and J. Y.T.~Wei$^1$}
\affiliation{$^1$Department of Physics, University of Toronto, 60 St. George Street, Toronto, ON M5S 1A7, Canada}
\affiliation{$^2$Department of Physics and Institute for Pure and Applied Physical Sciences, University of California San Diego, La Jolla, California 92093}

\begin{abstract}

Point-contact Andreev reflection spectroscopy was performed on single crystals of the heavy-fermion superconductor PrOs$_{4}$Sb$_{12}$, down to 90mK and up to 3Tesla. The conductance spectra showed multiple structures, including zero-bias peaks and spectral dips, which are interpreted as signatures of multiple pairing gaps, one clearly having nodes. Samples with 2\% Ru replacing Os were also measured, showing pronounced spectral humps consistent with the emergence of a non-nodal gap.  Detailed analysis of the spectral evolution revealed a magnetic field-\emph{vs.}-temperature phase diagram for PrOs$_{4}$Sb$_{12}$ characterized by two distinct superconducting order parameters.

\end{abstract}


\maketitle

Superconductivity in heavy-fermion materials has been a heavily researched topic, particularly for the role that strongly-correlated electrons play in the pairing process.  The discovery of superconductivity in the filled skutterudite PrOs$_4$Sb$_{12}$ has attracted much attention, because it is the first heavy-fermion superconductor containing neither Ce nor U but Pr atoms which show no ground-state magnetic order \cite{BauerPRB}.  Various unconventional properties have been reported in the superconducting state of PrOs$_4$Sb$_{12}$ \cite{MapleJPSJS}.  Of particular interest is the appearance of two superconducting transition temperatures ($T_{c}$'s) \cite{VollmerPRL,TayamaJPSJ} and low-energy quasiparticle excitations \cite{ChiaPRL,HuxleyPRL,Fredericknew}, both phenomena suggesting an exotic pairing mechanism.  Furthermore, angular magneto-thermal conductivity measurements have indicated that the pairing symmetry of PrOs$_4$Sb$_{12}$ undergoes a phase transition in a magnetic field \cite{IzawaPRL}.  Such a complex order-parameter (OP) phase diagram suggests the presence of multiple superconducting OPs, reminiscent of the case of UPt$_{3}$ \cite{JoyntReview}.  

There have been conflicting experimental reports on the superconducting gap topology of PrOs$_4$Sb$_{12}$, some indicating the presence of gap nodes while others indicating the Fermi surface to be fully gapped \cite{Maplereview}.  To reconcile the difference between these reports, a recent proposal has invoked the multi-band dispersion of PrOs$_4$Sb$_{12}$ \cite{SugawaraPRB} to suggest that there are two superconducting gaps, i.e. a nodal gap in a light-mass band and a non-nodal gap in a heavy-mass band \cite{MacPhysica, Shu_PRB09}.  Double superconducting gaps have been observed in two separate thermal-conductivity measurements \cite{SeyfarthPRL2,HillPRL}.  However, whereas the earlier measurement indicated both gaps to have $s$-wave symmetry \cite{SeyfarthPRL2}, the more recent measurement indicated only one $s$-wave gap with the other gap having nodal symmetry \cite{HillPRL}.  The latter result suggests that there are two distinct OPs with different pairing symmetries and $T_{c}$'s in PrOs$_4$Sb$_{12}$ \cite{note_mbsc}.  Although still under debate, this multi-band and multi-symmetry scenario could also explain the two $T_{c}$'s seen in heat-capacity measurements on Pr(Os$_{1-x}$Ru$_{x}$)$_{4}$Sb$_{12}$, where the nodal gap becomes dominated by a non-nodal gap for Ru-doping above $\approx$1\% \cite{Fredericknew}. 

Despite these prior studies, it is still not clear whether PrOs$_4$Sb$_{12}$ has one or more superconducting gaps, and whether these gaps have the same symmetry.  The vast majority of these studies have involved \emph{bulk} probes which could not directly detect the effects of sample inhomogeneity.  In this paper, we report on point-contact Andreev reflection spectroscopy (PCARS) measurements of PrOs$_4$Sb$_{12}$ single crystals. PCARS is an inherently \emph{local} probe and also sensitive to phase of the superconducting OP \cite{Deutscher:rev}.  It has been used to probe the pairing state of several unconventional superconductors, including cuprates \cite{Zad}, ruthenates \cite{LaubePRL, MaoPRL}, borocarbides \cite{BobrovPRB}, MgB$_{2}$ \cite{SzaboPRL} and heavy-fermion metals \cite{NaidyukRev,Goll:book}.  Our measurements on PrOs$_4$Sb$_{12}$ showed multiple spectral features which provide \emph{microscopic} evidence for two superconducting gaps, one clearly having nodes.  We also measured crystals with 2\% Ru-doping, which showed the emergence of a non-nodal superconducting gap.  Detailed analysis of the spectral evolution down to 90mK and up to 3Tesla yielded a magnetic field-\emph{vs.}-temperature phase diagram for PrOs$_4$Sb$_{12}$, characterized by two distinct superconducting OPs with different symmetries.  

The single crystals used in our experiment were grown with a molten metal-flux method \cite{BauerJPhys}.  For the undoped PrOs$_4$Sb$_{12}$ crystals, electrical resistivity $\rho $ measurements showed a single $T_{c}$ at $\approx1.85$K with $\approx5$mK transition width, and an upper critical field $H_{c2}\approx2.25$T at 200mK.  AC susceptibility $\chi (T)$ showed two $T_{c}$'s, i.e. $T_{c1}\approx1.85$K and $T_{c2}\approx1.65$K, similar to previously reported results \cite{VollmerPRL, GrubePRB}.  PCARS measurements were performed in a $^{3}$He-$^{4}$He dilution refrigerator, using Pt-Ir tips on \textit{c}-axis faces of the crystals.  The crystals were etched in a 1:1 HNO$_{3}$-HCl mixture to remove any residual Sb flux.  Junction impedances were typically 0.5-1 $\Omega$, resulting in point contacts which were in the ballistic regime \cite{note_ballistic}.  To minimize Joule heating, a pulsed-signal four-point technique \cite{RourkePRL} was used to acquire current-\emph{vs.}-voltage $I$-$V$ curves, which were numerically differentiated to yield $dI/dV$ spectra.

\begin{figure}[h!]
\centering
\includegraphics[width=3.2in]{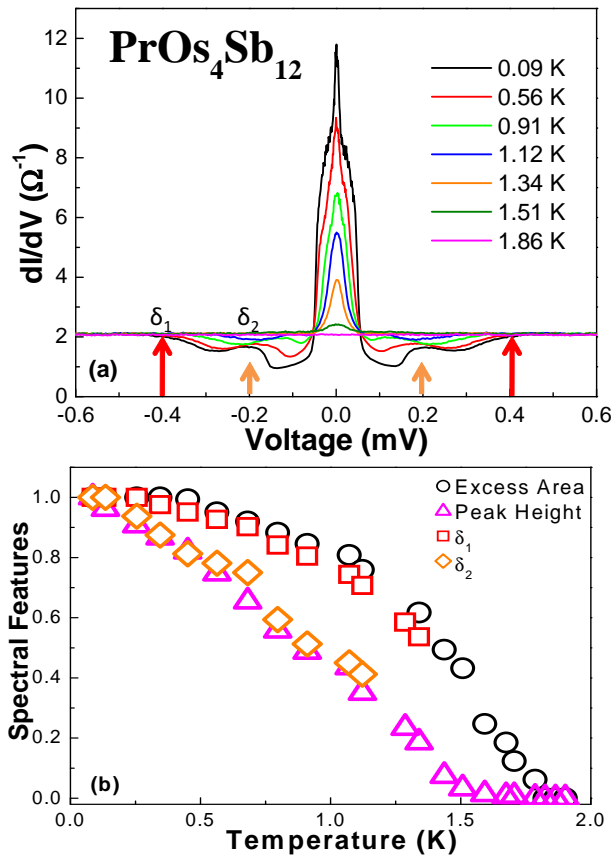}
\caption{(color online). (a) Differential conductance as a function temperature for a Pt-Ir/PrOs$_4$Sb$_{12}$ point-contact junction in zero magnetic field.  (b) Temperature dependence of various spectral features, after normalizing relative to their maximum values.  The triangles correspond to zero-bias peak height, the squares and diamonds to $\delta _{1}$ and $\delta _{2}$. The circles correspond to the excess spectral area, as defined in the text.}
\label{temp}
\end{figure}

Figure \ref{temp}(a) shows the temperature evolution of $dI/dV$ spectra taken on PrOs$_{4}$Sb$_{12}$ in zero magnetic field.  At 90mK, well below $T_{c1}\approx1.85$K, there is a pronounced zero-bias peak (ZBP) accompanied by a symmetric dip which begins at $\approx$$\pm$0.4mV, as indicated by $\delta _{1}$ in the figure.  An additional dip is also present at $\approx$$\pm$0.2mV, as indicated by $\delta _{2}$.  With increasing temperature, the ZBP lowers in height and the dips move inward.  The spectrum flattens out above $T_{c1}\approx1.85$K, consistent with the sample no longer being superconducting.  Figure \ref{temp}(b) shows detailed temperature evolution of the ZBP height and of both $\delta _{1}$ and $\delta _{2}$ positions, after normalizing them relative to their maximum values.  Also plotted in Fig.\ref{temp}(b) is the excess spectral area, which is defined by subtracting each $dI/dV$ spectrum by the normal-state spectrum and then numerically integrating between $\pm$ 1.8mV.  These plots indicate that the evolution of the various spectral features is governed by two distinct temperature scales, corresponding respectively to $T_{c1}\approx1.85$K and $T_{c2}\approx1.65$K.

The ZBPs seen in our $dI/dV$ spectra can be interpreted as distinctive signature of a \emph{nodal} gap, while the excess spectral area provides strong evidence for a \emph{non-nodal} second gap.  First, several bulk experiments have already indicated PrOs$_{4}$Sb$_{12}$ to have point nodes in its superconducting gap structure \cite{IzawaPRL, ChiaPRL, HuxleyPRL,Shu_PRB09}.  Second, theoretical modeling of Andreev reflection for PrOs$_{4}$Sb$_{12}$ has shown that points nodes in a triplet pair potential can produce ZBPs \cite{AsanoPOS,LinderPRB}, similar to the effect of $d$-wave line nodes for cuprates \cite{WeiPRL}.  Here we point out that ZBPs are a manifestation of nodal Andreev states, which are due to interference between quasiparticles  and thus conserved, in contrast to the non-nodal Andreev states, which are due to retroflection from pairs and thus not conserved \cite{BTK,RourkePRL}.  Therefore, the excess spectral area seen in Fig.\ref{temp} indicates that, in addition to contributions from Andreev states due to a \emph{nodal} gap, there are also Andreev states due to a \emph{non-nodal} gap \cite{note_BCS}.  In fact, since the ZBP vanishes near $T_{c2}\approx1.65$K while the excess area persists up to $T_{c1}\approx1.85$K, these spectral features can be attributed to two distinct superconducting OPs \cite{note_mbsc}.  In essence, our data provides microscopic corroboration for the thermal conductivity results of Ref.\cite{HillPRL}, indicating two superconducting gaps with different symmetries in PrOs$_{4}$Sb$_{12}$.

\begin{figure}[h!]
\centering
\includegraphics[width=3.2in]{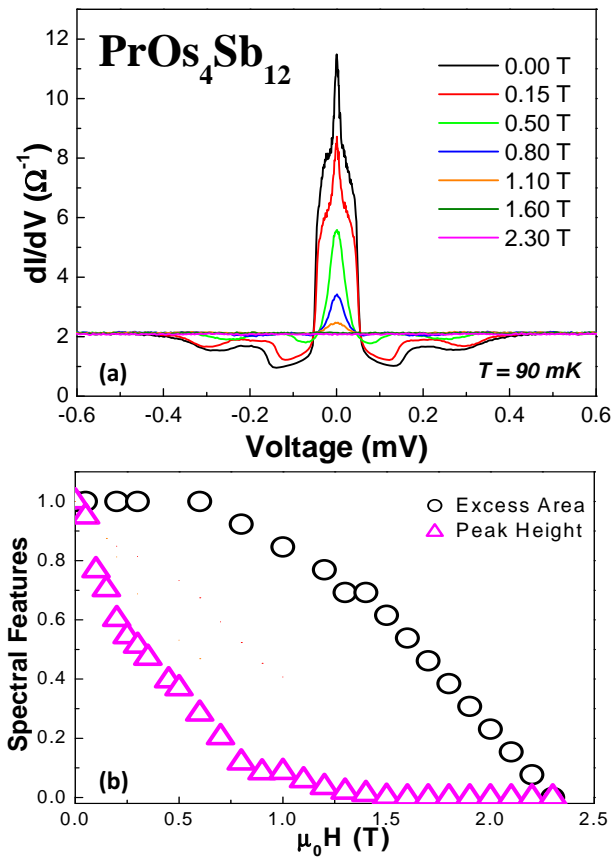}
\caption{(color online). (a) Differential conductance as a function of magnetic field for a Pt-Ir/PrOs$_4$Sb$_{12}$ point-contact junction at 90mK.  (b) Field dependence of the zero-bias peak height and excess spectral area, as defined in the text.}
\label{field}
\end{figure}

Figure \ref{field}(a) shows the magnetic-field evolution of $dI/dV$ spectra measured at 90mK.  As the field is increased, the ZBP lowers in height and the dip structures move inward, qualitatively similar to the spectral evolution with temperature.  However, the ZBP height collapses more quickly with field than with temperature, relative to the evolution of the excess spectral area.  As shown in Figure \ref{field}(b), the ZBP height vanishes above $\approx$1.5T while the excess area persists up to $\approx$2.3T.  Using the two-OP Andreev reflection scenario outlined above, this field evolution indicates that there are also two different energy scales governing the field suppression of superconductivity in PrOs$_4$Sb$_{12}$.  Namely, the \emph{nodal} Andreev states are spectrally robust up to a lower-field boundary $H'\approx1.5$T at 90mK, while the \emph{non-nodal} Andreev states persist up to a higher-field boundary $H''\approx2.3$T at 90mK.  This two-boundary field evolution is qualitatively similar to the angular magneto-thermal conductivity data of Ref.\cite{IzawaPRL}, which indicated a phase-diagram boundary $H^{*}(T)$ across which the pairing symmetry changes.  It is worth noting that PCARS data taken on UPt$_{3}$ \cite{GollPRB}, which is known to have multiple superconducting OPs \cite{JoyntReview}, also shows spectral features that vanish at a field boundary lower than $H_{c2}$.

\begin{figure}
\centering
\includegraphics[width=3.3in]{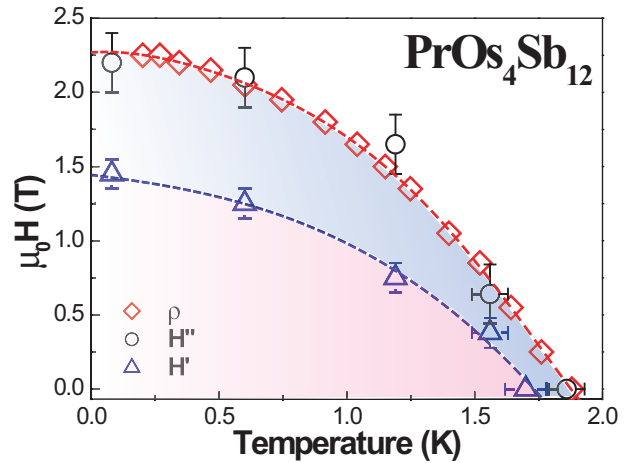}
\caption{(color online).  Magnetic field vs. temperature phase diagram determined from our point-contact spectroscopy and resistivity data on PrOs$_4$Sb$_{12}$.  $H'$ (triangles) correspond to the field at which the ZBP vanishes.  $H''$ (circles) correspond to the field at which the excess spectral area vanishes.  $H_{c2}$ (diamonds) is the upper-critical field measured by resistivity.  Dotted curves are added to guide the eye.  It should be emphasized that this phase diagram, which indicates two distinct superconducting OPs with different symmetries, is determined with a highly local probe, thus arguing strongly against sample inhomogeneity as the cause of multiple $T_{c}$'s.}
\label{phased}
\end{figure}

To map out a detailed $H$-$T$ phase diagram for PrOs$_4$Sb$_{12}$, we carried out a combined analysis of the temperature and field evolutions of our data.  Figure \ref{phased} shows a master plot of the full temperature dependences of $H'$ and $H''$ determined from our spectroscopy data, along with $H_{c2}(T)$ determined from our $\rho(T,H)$ data. First, it is clear that $H''(T)$ coincides with $H_{c2}(T)$, indicating that the higher-field boundary is just the resistive upper-critical field.  Second, $H''(T)$ and $H'(T)$ appear to emerge from different low-temperature asymptotes and gradually approach each other with increasing temperature.  This \emph{non-parallel} behavior between $H''(T)$ and $H'(T)$ resembles the $H$-$T$ phase diagram determined from angular magneto-thermal conductivity in Ref.\cite{IzawaPRL}, in contrast to \emph{parallel} phase boundaries determined from heat capacity in Ref.\cite{MeassonPRB}.  Finally, our $H''(T)$ and $H'(T)$ curves appear to approach zero at different temperatures,  $\approx$1.85K and $\approx$1.65K respectively, agreeing well with $T_{c1}$ and $T_{c2}$ measured by $\chi(T)$ on our samples.  Here it is important to emphasize that our $H$-$T$ diagram, which indicates two distinct superconducting OPs with different symmetries and $T_{c}$'s, is determined spectroscopically with a highly local probe, thus arguing strongly against sample inhomogeneity as the cause of multiple $T_{c}$'s.  It should also be noted that although our $H$-$T$ diagram is qualitatively similar with the one reported in Ref.\cite{IzawaPRL}, there is quantitative difference in the low-temperature asymptote between our $H'(T)$ and their lower-field boundary $H^{*}(T)$.  Interestingly, surface impedance measurements on PrOs$_4$Sb$_{12}$ have also indicated a lower-field phase boundary with a zero-temperature asymptote of $\approx$1.5T \cite{TouP}, which is quantitatively consistent with our $H'(T)$.

\begin{figure}
\centering
\includegraphics[width=3.2in]{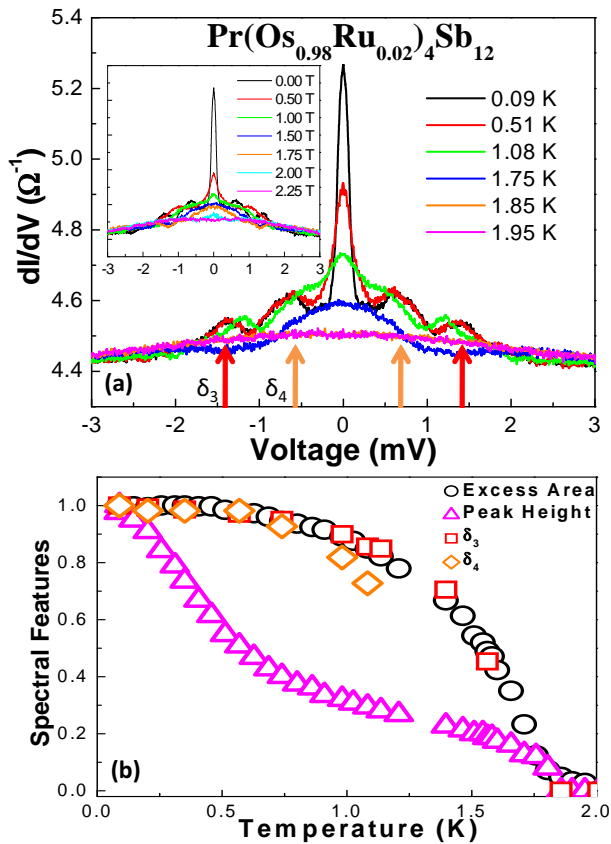}
\caption{(color online). (a) Differential conductance spectrum as a function of temperature for a Pt-Ir/Pr(Os$_{0.98}$Ru$_{0.02}$)$_{4}$Sb$_{12}$ point-contact junction in zero magnetic field.  The inset shows the spectral evolution versus field at 90 mK. (b) Temperature dependence of the ZBP height and excess spectral area, defined in the same way as in Fig.1.}
\label{dope}
\end{figure}

To probe the effects of Ru-doping on the two superconducting OPs observed, we also performed PCARS measurements on Pr(Os$_{0.98}$Ru$_{0.02}$)$_{4}$Sb$_{12}$ crystals. These 2\% Ru-doped crystals have $T_{c} \approx1.8$K and $H_{c2} \approx2.2$T at 200mK.  Figure \ref{dope}(a) plots the temperature evolution of a typical $dI/dV$ spectrum.  At 90mK a ZBP is also observed, but its height is lower than in the undoped case.  Broad hump structures clearly emerge at $\approx$$\pm$1.4mV and $\approx$$\pm$0.6mV as indicated by $\delta_{3}$ and $\delta_{4}$ respectively in the figure.  With increasing temperature, the ZBP decreases in height while the humps become narrower, and the spectrum completely flattens out above $T_{c}\approx1.8$K.  The inset of Fig.\ref{dope}(a) shows the spectral evolution with magnetic field at 90mK.  While the ZBP collapses into the hump structures above $\approx$1.5T, a spectral hump is still visible up to $\approx$2.2T, unlike the undoped case which shows no pronounced humps (Fig.\ref{field}(a)).  Figure \ref{dope}(b) plots the temperature dependences of the ZBP height, the excess spectral area, and the positions of $\delta _{3}$ and $\delta _{4}$, using the same data reduction as for Fig.\ref{temp}(b).  Comparing Fig.\ref{dope}(b) and Fig.\ref{temp}(b), it is clear that while both cases show excess spectral area up to the resistive $T_{c}$, Fig.\ref{dope}(b) shows a more rapid ZBP collapse with temperature.  Using the two-OP Andreev reflection scenario from above, these observations indicate that Ru-doping strengthens the \emph{non-nodal} OP while weakening the \emph{nodal} OP, in agreement with both heat-capacity and penetration-depth measurements \cite{Fredericknew,ChiaJP} which showed the rapid emergence of a \emph{non-nodal} OP with Ru-doping.   It is interesting to note that, for our PCARS data with 2\% Ru-doping, the higher of the two temperature scales governing the spectral evolution appears to correspond to two hump structures, denoted by $\delta _{3}$ and $\delta _{4}$.  These hump structures suggest that the \emph{non-nodal} OP itself may have multiple sub-components. Further measurements of crystals with higher Ru-doping levels are under way to elucidate this possibility.

In summary, we have performed point-contact Andreev reflection spectroscopy on PrOs$_4$Sb$_{12}$ down to 90mK and up to 3Tesla, and observed \emph{microscopic} evidence for multiple superconducting OPs.  Detailed analysis of the spectral evolution revealed a $H$-$T$ phase diagram characterized by two distinct OPs, one of which clearly having gap nodes and the other one being non-nodal and becoming more robust with just 2\% Ru-doping.  Our observations corroborate similar results from bulk measurements, and highlight the unconventional nature of the pairing state in this heavy fermion superconductor.

\begin{acknowledgments}
Research at the University of Toronto was supported by grants from NSERC, CFI/OIT, OCE, and the Canadian Institute for Advanced Research under the Quantum
Materials Program.  Research at UCSD was supported by the US Department of Energy under research grant number DE FG02-04ER46105.
\end{acknowledgments}


\begin{references}
\bibitem{note_mbsc} In the standard model of multiband superconductivity \cite{Suhl}, sub-components of a single OP are expected to share a common pairing symmetry and transition temperature.
\bibitem{note_ballistic} Using the Wexler formula, $R \approxeq 4\rho l/3\pi a^{2} + \rho /2a$, with $\rho$ $\approx$ 2 $\mu\Omega$cm, mean free path $l$ $\approx$ 350 nm, and $R$ the junction impedance, the contact radius, $a$ is estimated to be $\approx$ 90 nm thus satisfying the ballistic criterion $a < l$.
\bibitem{note_BCS} Note that the excess area has a BCS-like temperature dependence, consistent with a fully-gapped OP; while the ZBP height deviates from such BCS-like behavior, consistent with a more complex OP having gap nodes \cite{DeWildePRL}.
\end{references}
\end{document}